\documentclass[10pt]{article}

\newcommand{\ignore}[1]{}

\usepackage{cite}

\renewenvironment{abstract}{%
\begin{quote} \bf}
{\end{quote}}

\title{\Large\bf Impact of Boundaries on Fully Connected Random Geometric Networks} 

\author
{Justin Coon,$^{1}$ Carl P. Dettmann,$^{2}$ Orestis Georgiou$^{3}$\\
\normalsize{\em $^{1}$ Toshiba Telecommunications Research Laboratory, 32 Queen Square, Bristol BS1 4ND, United Kingdom}\\
\normalsize{\em $^{2}$ University of Bristol School of Mathematics, University Walk, Bristol BS8 1TW, United Kingdom}\\
\normalsize{\em $^{3}$ Max Planck Institute for the Physics of Complex Systems, N\"othnitzer Str. 38, 01187 Dresden, Germany}
}

\usepackage{graphicx}

\begin{document}
\maketitle

\begin{abstract}
Many complex networks exhibit a percolation transition involving a macroscopic connected
component, with universal features largely independent of the microscopic model and the
macroscopic domain geometry. In contrast, we show that the transition to full connectivity
is strongly influenced by details of
the boundary, but observe an alternative form of universality. Our approach correctly distinguishes
connectivity properties of networks in domains with equal bulk contributions.  It also facilitates
system design to promote or avoid full connectivity for diverse geometries in arbitrary dimension.
\end{abstract}

\maketitle

\section{Introduction}
Random geometric network models~\cite{Penrose, FM} comprise a collection of entities
called nodes embedded in region of typically two or three dimensions, together
with connecting links between pairs of nodes that exist with a probability related
to the node locations. They appear in numerous complex systems including
in nanoscience~\cite{Kyr}, epidemiology~\cite{Miller,Danon}, forest fires~\cite{Pueyo},
social networks~\cite{Palla,Parshani}, and wireless
communications~\cite{HABDF,Li,Wang}. Such networks exhibit a general
phenomenon called {\em percolation}~\cite{Callaway,BR}, where at a critical connection probability
(controlled by the node density), the largest connected
component (cluster) of the network jumps abruptly from being independent of system size (microscopic) to being proportional to system size (macroscopic).

Percolation phenomena are closely related to thermodynamic phase transitions where the number of nodes
$N$ goes to infinity and the critical percolation density $\rho_{c}$ is largely independent of the system size, shape,
and of the microscopic details of the model; the phenomenon of universality.
At the critical point, conformal invariance in two dimensional networks leads to detailed expressions for the probability
of a connection across general regions~\cite{Cardy} and more general connections with conformal field theory~\cite{Fuchs} and Schramm-Loewner Evolution~\cite{StAubin}.
Here, we take a different approach and are concerned with finite networks and with questions related to percolation,
but fundamentally different: What node density ensures a specified probability $P_{fc}$ that
the entire network is a single connected component (cluster), that is, {\em fully connected}?  How is this probability
affected by the shape of the network domain?

These questions are crucial for many applications, including for example the design of
reliable wireless mesh networks. These consist of communication devices
(the nodes) that pass messages to each other via other nodes rather
than a central router. This allows the network to operate seamlessly over a large
area, even when nodes are moved or deactivated. A fully connected network means that every node can communicate with every other node through direct or indirect connections. Mesh networks have been developed for many communication systems, including laptops, power distribution (``smart grid'') technologies, vehicles for road safety or environmental monitoring, and robots in hazardous locations such as factories, mines and disaster areas~\cite{Li}.

For many applications of random geometric networks including those above,
{\em direct connection} between two nodes $i$ and $j$ can be well described
by a probability $H_{ij}=H(r_{ij})$, a given function of the distance between the nodes
$r_{ij}=|{\bf r}_i-{\bf r}_j|$.
Often, the nodes are mobile or otherwise not located in advance, hence we assume $N$ uniformly distributed nodes confined in
a specified $d$-dimensional region $\cal V$ with area ($d=2$) or volume ($d=3$)
denoted by $V$. The node density is then defined as $\rho=N/V$.
For reference, we will later take $H(r_{ij})=\exp[-(r_{ij}/r_{0})^\eta]$, where $r_{0}$ is a
relevant length scale, and $\eta$ determines the sharpness
of the cut-off. Note that when $\eta\rightarrow\infty$ a step function corresponding to the popular
{\em unit disk} deterministic model~\cite{DJLN} is obtained, where connections have a
fixed range $r_0$. Our derivation however is completely general and only requires that
$H_{ij}$ is sufficiently short-ranged compared to system size.
Using this as a basis, we find that contrary to common belief and practice, the
geometrical details of the confined space boundaries ({\em corners, edges} and {\em faces}) dominate the
properties of the percolation transition. Moreover, the short-range nature of $H_{ij}$ allows us to separate individual boundary components and
obtain analytic expressions for $P_{fc}$ at high densities as a sum over their contributions. We confirm this through computer simulations and argue that the substantial improvement
offered by our main result Eq.~\ref{e:bdy} can be used to predict, control, optimize or even set benchmarks
for achieving full network connectivity in a wide variety of suitable models and applications involving finite size geometries.

\begin{figure}
\centerline{\includegraphics[width=450pt]{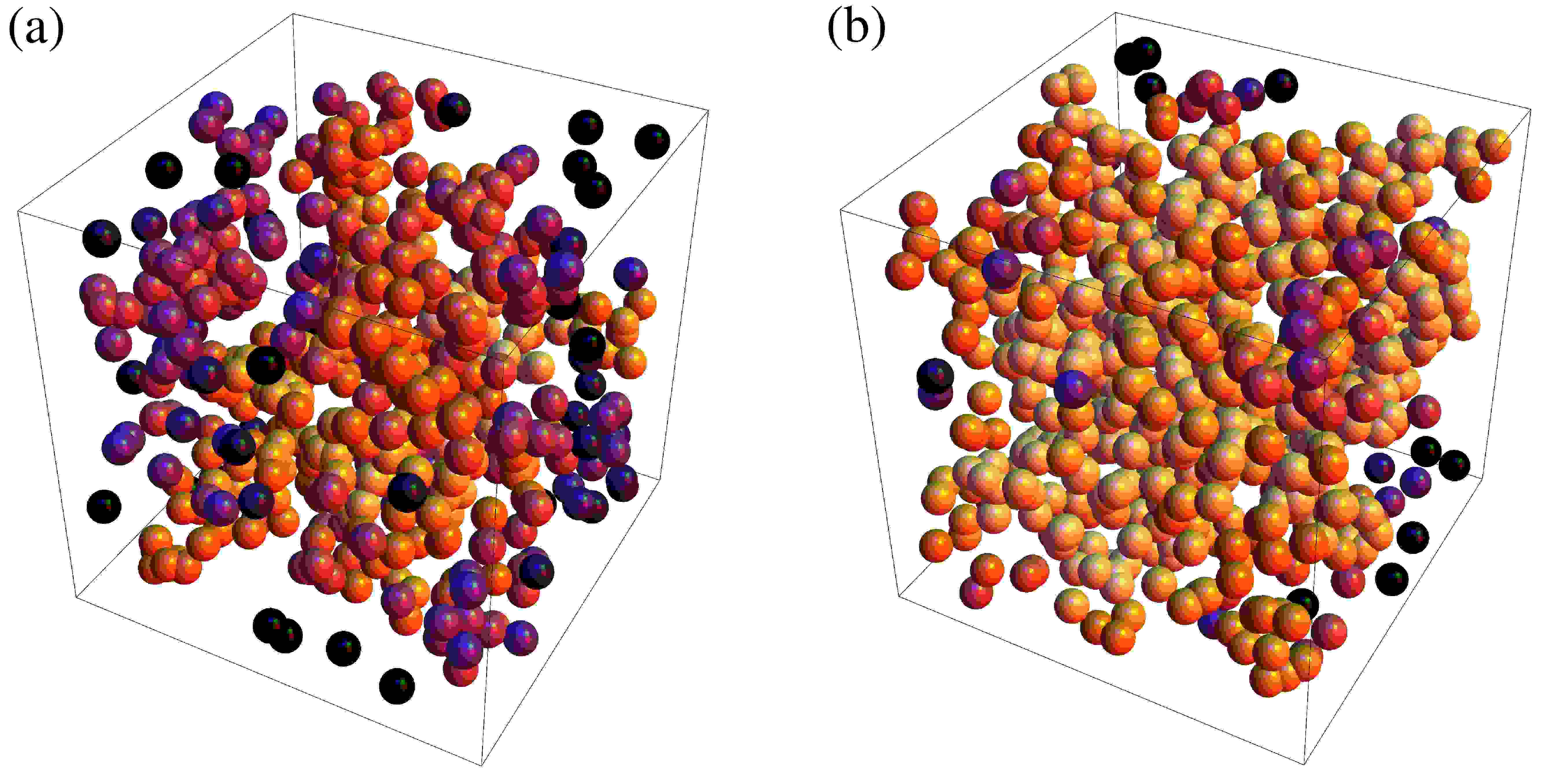}}
\caption{\label{f:balls}
Isolated nodes shown as black balls concentrate at the boundaries of the domain and particularly near corners at higher densities. Nodes are placed randomly in a cube, with lighter
colors indicating a higher probability of being in the largest connected component. We use $\eta=2$, while the side length of the cube is $L=10r_0$.
There are 500 nodes in (a) and 700 nodes in (b).}
\end{figure}

\section{Full connection probability}
As in conventional continuum percolation theory~\cite{Stell}, we start by utilizing a cluster
expansion approach~\cite{Hill} to derive a systematic perturbative method for
determining the full connection probability $P_{fc}$ as a function of density $\rho$.
Formulation of the expansion can be summarized as follows.
The probability of two nodes being connected (or not) leads to the trivial identity
$1\equiv H_{ij}+ (1-H_{ij})$.
Multiplying over all links expresses the probabilities $\mathcal{H}_g$ of all
$2^{N(N-1)/2}$ possible graphs $g$,
\begin{equation}
1= \prod_{i<j}[H_{ij}+ (1-H_{ij})]\equiv \sum_{g}\mathcal{H}_{g},
\end{equation}
Collecting terms according to largest cluster size we get
\begin{equation}\label{2}
1= \sum_{g\in \mathcal{G}_N}\mathcal{H}_{g} + \sum_{g\in \mathcal{G}_{N-1}}\mathcal{H}_{g} + \ldots + \sum_{g\in \mathcal{G}_{1}}\mathcal{H}_{g},
\end{equation}
where $\mathcal{G}_{n}$ is the set of all possible graphs with largest cluster of size
$n\in\{1\ldots N\}$. The first term on the right hand side is the probability of being fully
connected given a specific configuration of nodes. The average over all random
configurations $\langle\rangle\equiv V^{-N}\int_{\mathcal{V}} d^{N}{\bf r}$ of this
quantity is thus the overall probability of obtaining a fully connected network $P_{fc}$.
Moreover, the main idea conveyed by Eq.~(\ref{2}) is that at high densities, full connectivity is most likely to be broken by a single isolated node (the $\mathcal{G}_{N-1}$ term); this is sufficient
detail for most applications. Further corrections incorporate the probability of several isolated
single nodes and smaller clusters of nodes, for which a systematic expansion can be
developed~\cite{CDG11}.

Averaging Eq.~(\ref{2}) over all configurations and noting that to leading order the $N-1$ cluster
is fully connected, and that all nodes are identical, the first order approximation becomes
\begin{eqnarray}\label{first}
P_{fc}&\approx&1-\langle\sum_{g\in\mathcal{G}_{N-1}}\mathcal{H}_g\rangle\nonumber\\
&=&1-N\langle\prod_{j=2}^N(1-H_{1j})\rangle\\
&=&1-\rho\int_{\cal V}\left(1-\frac{M({\bf r}_1)}{V}
\right)^{N-1} {\rm d}{\bf r}_1\;\;,\nonumber
\end{eqnarray}
where the ``connectivity mass'' accessible from a node placed at ${\bf r}_1$ is given by
\begin{equation}\label{mass}
M({\bf r}_1)=\int_{\cal V} H(r_{12}){\rm d}{\bf r}_2 ,
\end{equation}
Assuming that the volume $V\gg \rho M({\bf r}_1)^2$ for any ${\bf r}_1$,
which is reasonable if the system is significantly larger than $r_{0}$ at moderate densities and that the number of nodes $N$ is large, Eq.\ref{first} simplifies
to
\begin{equation}\label{e:MA}
P_{fc}\approx 1-\rho\int_{\cal V} e^{-\rho M({\bf r}_1)}{\rm d}{\bf r}_1\;\;.
\end{equation}
This equation is equivalent to Eq.~(8) in Mao and Anderson~\cite{Mao} which
was derived for the specific case of a square domain. Following numerous
studies by probabilists and engineers~\cite{Penrose,FM}, these authors however
assumed an exponential scaling of system size $V$ with $\rho$ which essentially renders boundary effects negligible. Scaling the system in such a way is a common approach
as it corresponds to the limit of infinite density at fixed connection probability, however
in practice this limit is approached only for unphysically large volumes.
In contrast, we do not assume exponential growth of $V$, and also consider far more
general geometries in any dimension $d\geq 1$.

Without an exponentially growing volume $V$, the behavior of the
full connection probability at high densities is qualitatively different: It is
controlled by the exponential in Eq.~(\ref{e:MA}), and hence node positions
${\bf r}_1$ where the connectivity mass is small, that is,
near the boundary of $\cal V$.
Thus in contrast to the usual situation in statistical mechanics, the boundaries (and in particular corners) are important, and we will see they in fact dominate.
We illustrate this in Fig.~\ref{f:balls} where nodes are placed randomly inside a cube and an average over a large number of possible graphs gives the connectivity of each node. Notice that isolated and hard-to-connect nodes shown as dark balls concentrate near the boundaries of the domain and particularly near corners at higher densities.
This observation forms
the basis of our work, and has led to a radically different understanding
of connectivity in confined geometries which we now detail further. 

\section{Boundary effects}
The contributions to the integrals in Eq.~(\ref{e:MA}) come from ${\bf r}_1$
at {\em boundary components} $B\subset \cal V$ of dimension $d_B$, for example the bulk, the faces, and right angled edges and corners of a cube,
with $d_B=3,2,1$ and $0$ respectively. The short-range nature of $H_{ij}$ allows us to isolate each boundary component, whilst to leading order the connectivity mass splits into independent
radial and angular integrals, depending only on the local geometry of $B$ and hence
\begin{equation}\label{e:sgf}
M_B=M({\bf r}_B)=\omega_B\int_0^\infty H(r) r^{d-1}{\rm d}r\;\;,
\end{equation}
where $\omega_B$ is the angle ($d=2$) or solid angle ($d=3$) subtended by $B$. For example, if ${\bf r}_{B}$ is near a corner of the cube then $\omega_B= (4\pi)/8$, while near an edge $\omega_B= (4\pi)/4$, near faces $\omega_B= (4\pi)/2$ and $\omega_B= (4\pi)$ for the bulk interior. Hence, from Eq.~(\ref{e:MA}) we see that corner contributions to $P_{fc}$ as a function of $\rho$ are exponentially larger than edge contributions which are themselves exponentially larger than face contributions etc. This simple argument shows that the dominant contribution to $P_{fc}$ at high densities comes from the ``pointiest'' corners.

Expanding $H(r_{12})$ about ${\bf r}_{2}$ near the corresponding boundary component we obtain a next to leading order expansion for $M({\bf r}_{B})$ which we can then use to approximately evaluate the integral in Eq. (\ref{e:MA}). Ignoring exponentially smaller correction terms and combining all boundary contributions we arrive at
our main result
\begin{equation}\label{e:bdy}
P_{fc}\approx 1-\rho\sum_BG_BV_B e^{-\rho M_B}\;\;,
\end{equation}
where  $V_B$ is the $d_B$-dimensional ``volume'' of each component (equal to one in the case of a $0$-dimensional corner and $V$ when $d_B=d$), $G_B$ is a geometrical factor depending on $B$ and implicitly on $H$ and $M_B$ is as in (\ref{e:sgf}); see examples below.
Notice that Eq.~(\ref{e:bdy}) is completely general as we have only assumed a short-ranged $H_{ij}$ and not used its specific form. Moreover, it also does not depend on using Euclidean distance and holds in any dimension $d\geq1$ and geometry where the lack of connectivity is dominated by a situation involving an $N-1$ cluster and a single disconnected node. Hence Eq.~(\ref{e:bdy}) is a powerful and useful multi-purpose tool 
for analyzing full network connectivity at high densities in a wide variety of suitable models and applications involving finite size geometries.

For example, in the context of single input single output (SISO) wireless communication
channels and a Rayleigh fading model \cite{TV}, information theory predicts
$H(r_{ij})=\exp[-(r_{ij}/r_0)^\eta]$ with $\eta$ an environment and wavelength dependent decay parameter equal to $2$ for free propagation, increasing to $\eta\approx 4$ for a cluttered environment, while $r_{0}$ depends on the minimum outage rate threshold.
For nodes confined to a cube of side length $L$ and $\eta=2$ we find $V_B=L^{d_B}$, $G_B=(2^{3-d_B-1}/\pi\rho r_0^2)^{3-d_B}$, and $M_B=(r_0\sqrt{\pi})^32^{d_B-3}$ with contributions from each of the eight corners, twelve edges, six faces and bulk. However the derivation is general: Once $G_B$ and $M_B$ have been evaluated for these boundary components (right angled edges etc.) by standard asymptotic analysis of the relevant
integrals, they apply to any geometry with these features and length scales significantly
larger than $r_0$. This independence on the large scale geometry
follows from the short-range nature of $H_{ij}$ and is a type of universality allowing for the calculation of $P_{fc}$ in complex high dimensional geometries without increased difficulty.

\begin{figure}
\centerline{\includegraphics[width=450pt]{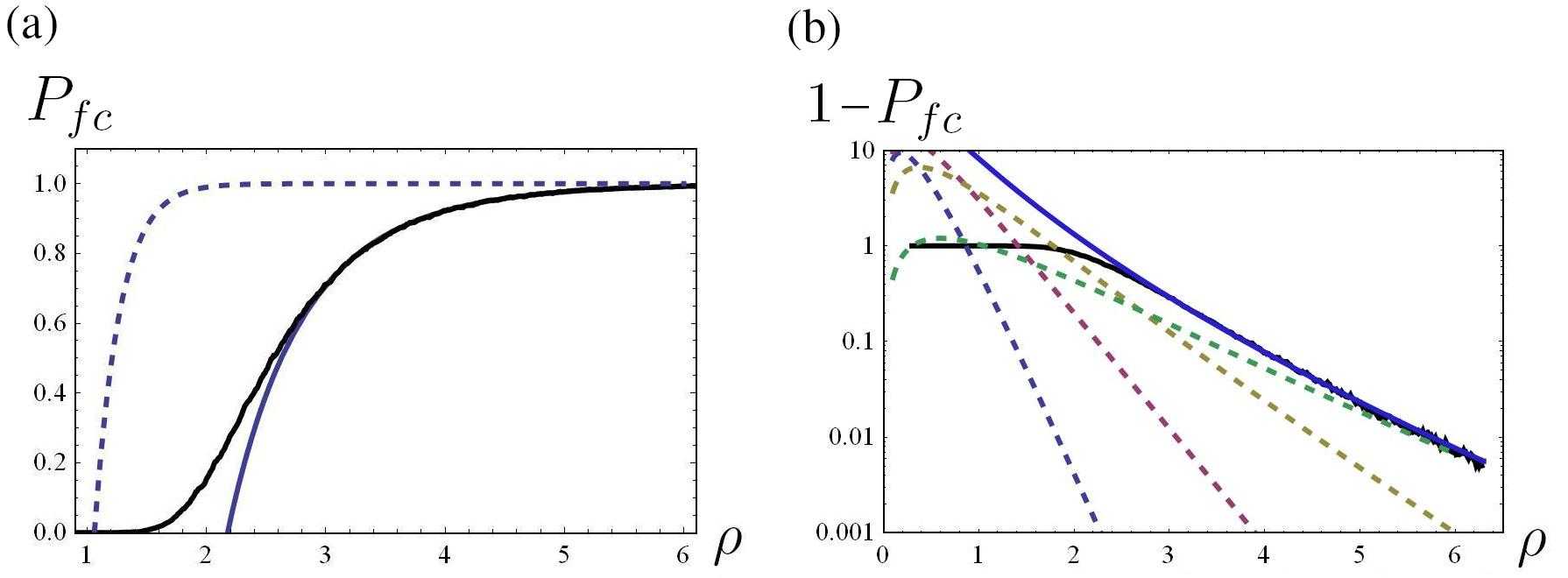}}
\caption{\label{f:cube}
(a)  Comparison of the full analytic prediction of Eq.~(\protect\ref{e:bdy}) (solid curve) with direct  numerical
simulation of the random network in a cube of side length $7r_0$ (jagged curve). The dashed line corresponds to the bulk contribution (previous theory).
(b) Contributions from the bulk (dotted blue, left), faces (red), edges (yellow) and
corners (green, right), together with the total (solid blue) and numerical simulation
(black jagged curve), showing the dominance of the corners at the highest densities and
good agreement between theory and simulation at moderate
to high densities.  Here it is convenient to plot
the outage probability $P_{out}=1-P_{fc}$.}
\end{figure}

The substantial improvement offered by Eq.~(\ref{e:bdy}) becomes clear when compared with the ``bulk'' contribution corresponding to current conventional wisdom shown in Fig.~\ref{f:cube}a) for a network confined to a cube. Fig.~\ref{f:cube}b) further demonstrates the inaccuracy of the bulk model as well as the benefits of including boundary effects when analyzing network connectivity in confined geometries.

\begin{figure}
\centerline{\includegraphics[width=450pt]{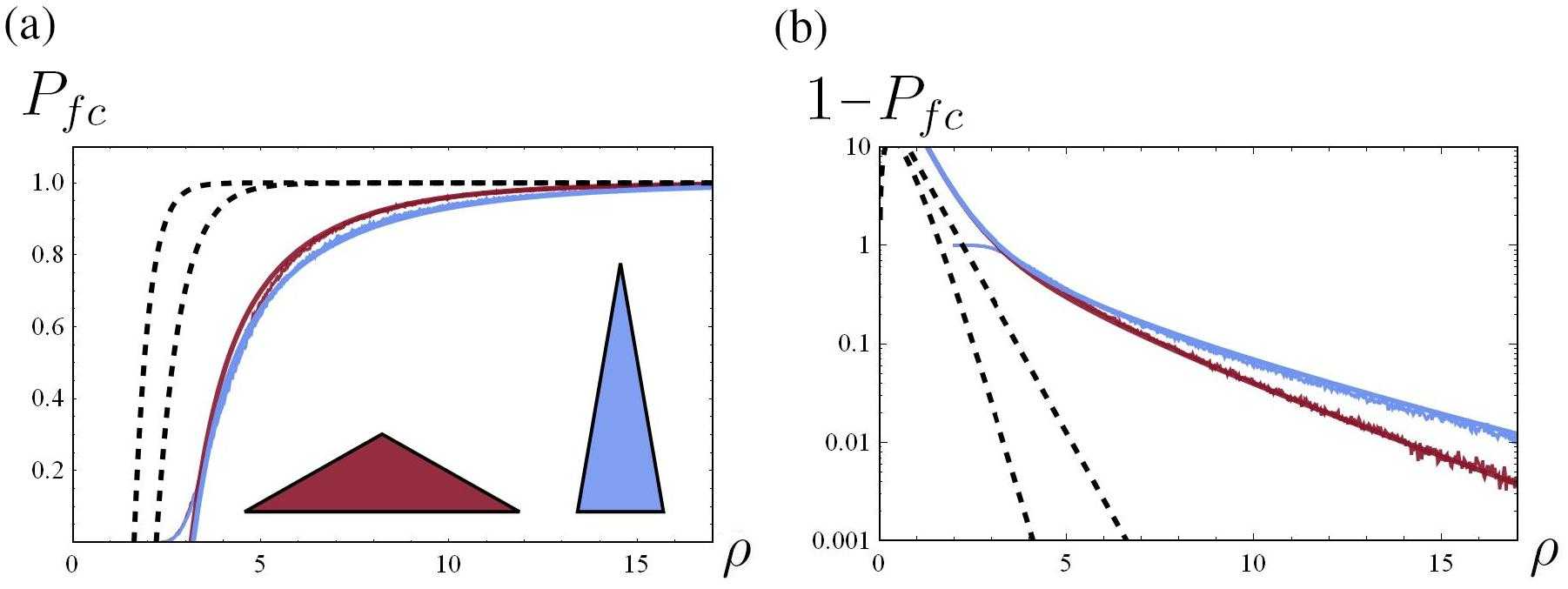}}
\caption{\label{f:tri}
Corner contributions in triangles with equal area and perimeter:
Comparison of theory with direct simulation, as in Fig.~\protect\ref{f:cube}.
The red triangle has side lengths of 26.88, 15.44 and 15.44 in units of the
connectivity length scale $r_0$, while the
blue triangle has side lengths of 8.40, 24.68 and 24.68.
The black dashed lines correspond to the equal bulk (left curve) and
bulk$+$edge (right curve) contributions while neglecting corner contributions. The colored curves give the
total (including crucial corner) contributions for each triangle. Both theory
and simulation are plotted, showing excellent agreement with the numerical simulations (jagged curves) which cover them completely for $\rho>4$.}
\end{figure}

We can go beyond simple geometries restricted to right-angled corners. Consider the case
of a two dimensional triangle
with general angles $0<\omega_B<\pi$. The relevant
integrals for this case come to $M_B=r_0^2\omega_B/2$, with $G_B=4/\pi\rho^2r_0^2\sin\omega_B$
for the corners and $G_B=(2^{2-d_B-1}/\pi\rho r_0^2)^{2-d_B}$ for the edges and bulk and can be generalized easily to higher dimensions.
Fig.~\ref{f:tri} shows two triangles chosen to have
identical perimeter and area; the connectivity at a given
density differs only due to the corner angles and agrees
perfectly with the full theory of Eq.~(\protect\ref{e:bdy}). A bulk theory,
even supplemented with edge contributions, is clearly
incapable of explaining the difference between the
connectivities of networks in these two triangles.
Moreover, such a situation motivates inverse problems, similar to ``hearing the shape of a drum" \cite{Kac66} by attempting to determine the size and shape details of an unknown domain containing a random network.

\section{Discussion}
An important aspect of the theory presented here is how it affects the design of real life random geometric networks.
For wireless mesh networks, the lack of connectivity near the boundaries can be mitigated
by increasing the signal power, the number of spatial channels, or by constructing a hybrid network with
a regular array of fixed nodes along the boundaries as well as randomly placed nodes in the interior. In each of these cases, the
design can now be analyzed given information about the cost and
connectivity function $H(r)$ and of course the desired connectivity region.
Conversely, boundary effects can be harnessed to avoid full connectivity if desired. For example in the case of forest fires \cite{Pueyo}
we have a prediction for the number of unburnt regions as a function of the geometric landscape and environment
parameters (for example angles between fire-lanes and/or natural boundaries), again given a specific model for
connectivity that depends on the type of vegetation, temperature, moisture content etc. Similar models could be
devised for the spread of epidemics \cite{Miller} or mobile phone viruses \cite{Wang} where boundaries are embedded in a more complex (possibly non-Euclidean) space yet
$H_{ij}$ is still short-ranged.

We examined connectivity in confined geometries and illustrated the importance of the often neglected boundary effects.
We then derived a general high density expansion Eq.~(\ref{e:bdy}) for the probability of full connectivity $P_{fc}$ assuming only a short-ranged connectivity function relative to system size and
showed that it displays universal features allowing for its easy calculation in complex geometries. This we have confirmed through computer simulations and argued that our approach
is well placed to facilitate efficiency in design in a variety of physical applications ranging from wireless networks to forest fire-lanes. Appropriate modifications of our theory can aid the understanding of small boundary-dominated systems such as for example the electrical conduction through carbon nanotubes in a
polymer matrix~\cite{Kyr} but possibly larger systems such as highly connected social and financial networks~\cite{Palla,Parshani}.

\section*{Acknowledgments}
The authors thank the Directors of the Toshiba Telecommunications Research Laboratory for
their support, and Charo Del Genio, Jon Keating and Mark Walters for helpful discussions.

\end{document}